\documentstyle[aps,preprint,tighten,psfig]{revtex}

\newcommand{\lsim}{\mathrel{\mathop{\kern 0pt \rlap
  {\raise.2ex\hbox{$<$}}}
  \lower.9ex\hbox{\kern-.190em $\sim$}}}
\newcommand{\gsim}{\mathrel{\mathop{\kern 0pt \rlap
  {\raise.2ex\hbox{$>$}}}
  \lower.9ex\hbox{\kern-.190em $\sim$}}}
\newcommand{\beq}    {\begin{equation}}
\newcommand{\eeq}    {\end{equation}}
\newcommand{\beqarr} {\begin{eqnarray}}
\newcommand{\eeqarr} {\end{eqnarray}}
\newcommand{\barr}   {\begin{array}}
\newcommand{\earr}   {\end{array}}

\begin{document}

\preprint{
\begin{tabular}{r}
DFTT 1/2000 \\
IFIC/00--12, FTUV/00--11 \\
LAPTH -- 779 /2000
\end{tabular}
}

\title{Further investigation of a relic neutralino
 as a possible origin of an annual--modulation effect in WIMP 
                        direct search}

\author{\bf 
A. Bottino$^{\mbox{a}}$
\footnote{E--mail: bottino@to.infn.it, donato@lapp.in2p3.fr, 
fornengo@flamenco.ific.uv.es, \\
\phantom{E--mail:~~~} scopel@posta.unizar.es},
F. Donato$^{\mbox{b}}$\footnote[4]{INFN Post--doctoral Fellow}, 
N. Fornengo$^{\mbox{a,c}}$,
S. Scopel$^{\mbox{d}}$
\vspace{6mm}
}

\address{
\begin{tabular}{c}
$^{\mbox{a}}$
Dipartimento di Fisica Teorica, Universit\`a di Torino \\
and INFN, Sez. di Torino, Via P. Giuria 1, I--10125 Torino, Italy\\
$^{\mbox{b}}$
Laboratoire de Physique  Th\'eorique LAPTH, B.P. 110, F--74941\\
Annecy--le--Vieux Cedex, France \\
$^{\mbox{c}}$
Instituto de F\'{\i}sica Corpuscular - C.S.I.C.\\
Departamento de F\'{\i}sica Te\`orica, Universitat de Val\`encia \\
Edificio Institutos de Paterna, Apt 2085, 46071 Valencia, Spain \\
$^{\mbox{d}}$ 
Instituto de F\'\i sica Nuclear y Altas Energ\'\i as, 
Facultad de Ciencias, \\
Universidad de Zaragoza, Plaza de San Francisco s/n, E--50009 Zaragoza, Spain
\end{tabular}
}

\maketitle

\begin{abstract}

We analyze the annual--modulation effect, measured by the DAMA 
Collaboration with the new implementation of a further two--years 
running,  in the context of a possible interpretation in  
terms of relic neutralinos. We impose over 
the set of supersymmetric configurations, selected by 
the annual--modulation data, the constraints derived from WIMP indirect 
measurements, and discuss the features of the ensuing relic 
neutralinos. We critically discuss the sources of the main 
 theoretical uncertainties in the analysis of event rates for 
direct and indirect WIMP searches.  
 
\end{abstract}  

\vspace{1cm}

\pacs{11.30.Pb,12.60.Jv,95.35.+d}

\section{Introduction}

  The effect of annual modulation measured by the DAMA 
Collaboration in its WIMP direct search experiment with 
a NaI(Tl) detector and reported in Ref. \cite{dama12} was 
analyzed  in terms of relic neutralinos in 
Refs.\cite{noi,noi5,noi6}. In these papers, 
  we proved  that this interpretation is compatible with the DAMA 
data, and entails  a relic neutralino which might have the role of 
 a major component of dark matter in the Universe, especially when  
the uncertainties affecting the evaluation of the neutralino--nucleon 
cross section are taken into account \cite{noi6}). We have also 
presented in detail other physical properties of such a neutralino, 
both in a Minimal Supersymmetric extension of the Standard Model 
\cite{noi,noi5,noi6} 
and in supergravity schemes \cite{noi}, and we have outlined how indirect 
measurements of WIMPS (low--energy antiprotons in cosmic rays and 
up-going muon fluxes from the center of the Earth and from the Sun) 
may  bring further information \cite{noi}, by way of 
constraints on the supersymmetric configurations derived from the 
DAMA annual--modulation results \cite{an}. 

 New data, collected by the DAMA Collaboration in a further 
two--year running of the NaI(Tl) experiment for an exposure of 
38 475 kg $\cdot$ day,  and now presented in 
Ref. \cite{dama3}, confirm their previous finding of an annual--modulation 
effect, which  does not appear to be related to any possible 
source of  random systematics. 
  Taking together all (old and new) samples of data for a total 
exposure of 57 986 kg $\cdot$ day, the effect turns out to be at a 
4$\sigma$ C.L. 
 Performing a maximum likelihood analysis in terms of 
$m_{\chi}$ and $\xi \, \sigma^{\rm (nucleon)}_{\rm scalar}$, where 
$m_{\chi}$ is the WIMP mass, 
$\sigma^{\rm (nucleon)}_{\rm scalar}$ is the WIMP--nucleon 
scalar elastic cross--section, and $\xi = \rho_{\chi}/{\rho_l}$ 
is the WIMP fractional amount of local non-baryonic  
dark--matter density $\rho_l$, the DAMA Collaboration presents a 
3$\sigma$ C.L.  annual--modulation region, in the plane 
$m_{\chi}$--$\xi \, \sigma^{\rm (nucleon)}_{\rm scalar}$, 
whose actual size depends on whether or not 
the upper--bound constraints previously obtained by the 
same Collaboration \cite{dama0} are included, and on the values assigned to
the galactic astrophysical velocities. 
For the purpose of the analysis carried out in the present paper, among the 
regions presented in Ref. \cite{dama3} we select the one, which is obtained 
from the annual--modulation data, by including the upper-bound 
constraints of Ref. \cite{dama0},  
by setting $\rho_l$ at the standard reference value: 
$\rho_l = 0.3$ GeV cm$^{-3}$, and by taking 
into account uncertainties in the astrophysical velocities of the usual
galactic Maxwellian distribution 
(170 km s$^{-1} \leq v_0 \leq$ 
270 km s$^{-1}$; $v_{esc}$ = 450--650 km s$^{-1}$; where $v_0$ is the 
rotational
velocity of the local system at the position of the solar system and 
$v_{esc}$ is the galactic escape velocity). 
This region is the one shown in Fig. 1 
(should one include also a bulk rotation of the dark matter halo 
\cite{dfs,noi5}, this region would elongate along the horizontal axis up to 
$m_{\chi} \sim$ 230 GeV \cite{dama3}). In this figure we also show the contour
lines for the three values $v_0 = 170, 220, 270$ km s$^{-1}$, separately 
\cite{nota3}. 
In the comparison of the experimental data with the theoretical evaluations 
one has to further consider the uncertainty in $\rho_l$:
 0.1 GeV cm$^{-3} \leq \rho_l \leq 0.7$ GeV cm$^{-3}$ \cite{turner1,turner}. 
Fig. 2 displays how the DAMA annual--modulation region shifts along the
vertical axis, as the value of $\rho_l$ is varied within its uncertainty range.
The four panels correspond to the representative values: 
$\rho_l$ = 0.1, 0.3, 0.5, 0.7 GeV cm$^{-3}$. In Fig. 2, as well as in all
subsequent figures, where experimental results of direct and indirect
WIMP measurements are compared with theoretical evaluations, separate panels
are used for the four representative values of $\rho_l$. 

    In the present paper we investigate the implications of the DAMA data with 
the total exposure of 57 986 kg $\cdot$ day   in terms of relic neutralinos, 
along the lines previously  developed in Refs.
\cite{noi,noi5,noi6}.
  We single out the set of the supersymmetric configurations 
compatible with the DAMA data, 
and then apply to this set, denoted as set $S$,  
the constraints derived from  experimental 
indirect searches for WIMPs (up-going muons 
at neutrino telescopes and antiprotons in cosmic rays). In this 
analysis we incorporate  recent, and quite significant,  
theoretical and experimental developments.

The supersymmetric theoretical framework adopted here is the Minimal 
Supersymmetric extension of the Standard Model (MSSM) 
\cite{susy}, which conveniently  describes the 
supersymmetric phenomenology at the electroweak scale, without too strong 
theoretical assumptions. This model has been extensively used by a number of
authors for evaluations of the neutralino relic abundance and detection rates
 (a list of references may be found, for instance, in  
\cite{bf}). 

The neutralino is defined 
as the lowest--mass linear superposition of photino ($\tilde \gamma$),
zino ($\tilde Z$) and the two higgsino states
($\tilde H_1^{\circ}$, $\tilde H_2^{\circ}$):
$\chi \equiv a_1 \tilde \gamma + a_2 \tilde Z + a_3 \tilde H_1^{\circ}  
+ a_4 \tilde H_2^{\circ}$. 

The MSSM contains three neutral Higgs fields: two of them 
($h$, $H$) 
are scalar and one ($A)$ is pseudoscalar. 
At the tree level the Higgs sector is specified by two independent parameters:
the mass of one of the physical Higgs fields, which we choose to
be the mass $m_A$ of the neutral pseudoscalar boson, and the ratio of the 
two vacuum expectation values, defined as $\tan\beta\equiv \langle H_2
\rangle/\langle H_1\rangle$. 
Once radiative corrections are introduced, the Higgs sector depends
also on the squark masses through loop diagrams. The radiative corrections 
to the neutral and charged Higgs bosons, employed in the present paper, are 
taken from Refs. \cite{radhiggs}.

\noindent
The other parameters of the model are defined in the superpotential, 
which contains all the Yukawa interactions
and the Higgs--mixing term 
$\mu H_1 H_2$, and  in the soft--breaking
Lagrangian, which contains the trilinear and bilinear  breaking 
parameters and the soft gaugino and scalar mass terms. 

To cast the MSSM, which originally contains a large number of parameters,  
into a form adequate for phenomenology, 
we follow the common procedure of introducing a set of 
restrictive assumptions at the electroweak scale: 
a) all trilinear parameters are set to zero except those of the third family, 
which are unified to a common value $A$;
b) all squarks and sleptons soft--mass parameters are taken as 
degenerate: $m_{\tilde l_i} = m_{\tilde q_i} \equiv m_0$, 
c) the gaugino masses are assumed to unify at $M_{GUT}$, and this implies that
the $U(1)$ and $SU(2)$ gaugino masses are related at the electroweak scale by 
$M_1= (5/3) \tan^2 \theta_W M_2$. 

Once these conditions are implemented in the model, the supersymmetric 
parameter space consists of six independent parameters. We choose them to be: 
$M_2, \mu, \tan\beta, m_A, m_0, A$ and vary these parameters in
the following ranges: $10\;\mbox{GeV} \leq M_2 \leq  1\;\mbox{TeV},\;
10\;\mbox{GeV} \leq |\mu| \leq  1\;\mbox{\rm TeV},\;
80\;\mbox{GeV} \leq m_A \leq  1\;\mbox{TeV},\; 
100\;\mbox{GeV} \leq m_0 \leq  1\;\mbox{TeV},\;
-3 \leq A \leq +3,\;
1 \leq \tan \beta \leq 50$. 
We remark that the values taken here as upper limits of the ranges for 
the dimensional parameters, $M_2, \mu, m_0, m_A$, are inspired by the upper 
bounds which may be
derived for these quantities in SUGRA theories, when one requires that the 
electroweak symmetry breaking, radiatively induced by the soft supersymmetry
breaking, does not occur with excessive fine tuning 
(see Ref. \cite{bbefms1} and references quoted therein).

We have further constrained our parameter space, by taking into account 
all the new experimental limits obtained from accelerators on
supersymmetric and Higgs searches (LEP2 \cite{lep2}, CDF \cite{cdf}). 
Notice that the new bounds from LEP2 and CDF constrain now rather severely the 
susy space, especially in the region of interest for direct detection 
(small $m_h$ and, partially, large $\tan \beta$ \cite{cdf}).

Moreover, the constraints 
due to the $b \rightarrow s + \gamma$ process \cite{bertolini} 
have been taken into
account. In our analysis, the inclusive decay rate 
BR($B \rightarrow X_s \gamma$) is calculated with corrections up to the 
leading order. Next--to--leading order corrections 
\cite{chetyrkin}  
are included only when 
they can be applied in a consistent way, i.e. both to standard--model 
and  to susy diagrams. 
We require that our theoretical evaluation for BR($B \rightarrow X_s \gamma$) 
is within the  range:  
1.96 $\times 10^{-4} \leq$ BR($B \rightarrow X_s \gamma$) $\leq$ 4.32
$\times 10^{-4}$. This range is obtained by combining the experimental 
data of Refs. \cite{cleo,barate} at 95\% C.L. and by adding a 
theoretical uncertainty of 25\%, whenever the still incomplete 
next--to--leading order susy corrections cannot be applied.

Our parameter space has been further constrained by the request that  
the Lightest Supersymmetric Particle (LSP) is the neutralino, rather than 
the gluino or squarks or sleptons. 
The current upper bound for cold dark matter may be establish as 
$\Omega_{CDM}h^2 \lsim 0.3$
($h$ is the usual Hubble parameter, defined in terms of the present--day 
value $H_0$ of the Hubble constant as 
$h \equiv H_0/(100~$ km$~$s$^{-1}~$Mpc$^{-1})$), on the basis of the most 
recent cosmological data 
\cite{cosmo}. However, for sake of presentation of the results of the present
analysis, which, anyway, never entail values of $\Omega_{\chi}h^2$ in excess of
0.6 (see last section), we do not impose the bound $\Omega_{CDM}h^2 \leq 0.3$ 
in our selection of susy configurations. 
The neutralino relic abundance is calculated here as illustrated in
Ref.\cite{ouromega}. 
We have checked that susy configurations which could potentially
lead to coannihilation effects \cite{coannih} are marginal in
our selected supersymmetric parameter space.

A few comments are in order here. The  restrictive  assumptions 
a) -- c) adopted above in the framework of the MSSM are instrumental in  
reducing 
the otherwise large number of independent parameters to a handful set of 
them (six in our scheme), and in making the calculations of a number of 
crucial observables (such as relic abundances and event rates) manageable. 
The few independent parameters of this simplified MSSM  have 
the role of relevant scales for some fundamental quantities, such as 
scalar masses and gaugino masses, which in turn determine the size of the 
numerical outputs.  This version of MSSM is obviously the simplest 
scheme for a susy model, and the most natural one to start with. However, 
one has to be aware of the fact that new experimental data could eventually 
force one to adopt more involved versions of supersymmetric models, for 
instance by relaxing some GUT-inspired relation (such as 
$M_1 \simeq 0.5 M_2$) Ref. \cite{m1m2}, or by including CP--violating 
phases \cite{cp}. 

 As regards the distribution of relic neutralinos in our Galaxy, 
to start with we have assumed  a standard halo population with a Maxwellian 
velocity distribution, whose dispersion speed is centered around 
270 km s$^{-1}$ (i.e., $v_0$ = 220 km s$^{-1}$). However, in the 
implementation of constraints from  
up-going muons at neutrino telescopes, we have also considered recent 
theoretical developments which may have quite contrasting effects on the 
expected signals \cite{dk,gould99}. 
 These different instances are examined in Sect. III. 

Data on antiprotons in space, combined with recent evaluations 
of the secondary antiproton component in cosmic rays due to spallation processes, 
are employed in Sect. IV to put further constraints on the original 
set $S$ of susy configurations, singled out by the DAMA data.   
  
We give the results of our combination of the annual--modulation data 
with indirect measurement constraints in Sect. V, where 
we also discuss the  cosmological properties for our set of relic neutralinos 
and present our conclusions.

\section{Set of supersymmetric configurations singled out by the 
annual--modulation data}

  In our papers of Ref. \cite{noi,noi5,noi6} 
we proved that the DAMA annual--modulation region of 
Ref. \cite{dama12} is widely compatible with an interpretation in terms of 
relic neutralinos, by showing that a sizeable portion of that region 
is covered by supersymmetric configurations, satisfying all accelerator 
bounds. Now we show in Fig. 1 that the new,  
more constrained annual--modulation region of Ref. \cite{dama3} 
is still largely compatible with the relic neutralino interpretation, 
though the 
supersymmetric space is now more severely constrained by the current 
limits from accelerators \cite{lep2,cdf}. 

In deriving the scatter plot shown in Fig. 1 we have used the scan of the 
susy parameter space  defined in the previous section. The 
neutralino--nucleon cross section has been 
calculated with the formulae reported in Ref. \cite{noi}. As discussed 
in Ref. \cite{noi6}, 
this cross section suffers from significant  
uncertainties in the size of Higgs--quark--quark and 
squark--quark--neutralino couplings. In fact, these couplings  
depend on quark masses $m_q$ and quark scalar densities in the nucleon 
$\langle \bar{q} q \rangle$, 
which are still rather poorly determined. To be specific, we refer to the
following quantities: 
 the fractional strange--quark content of the nucleon 
$y=2 \; <\bar ss> / (<\bar uu+ \bar dd>)$,
  the quark mass
ratio $r=2m_s/(m_u+m_d)$, and the products $m_{q}<\bar{q}q>$'s. 
 In our analysis we have  taken into account 
the uncertainties in these quantities. 
Thus, our scatter plots comprise representative points which have
been derived by using both of the two following sets of 
values, cumulatively:

\begin{eqnarray} \label{eq:set1}
& \mbox{Set 1:}\phantom{setset} & y = 0.33, \; r = 29,\nonumber \\
& & m_{l}<\bar{l}l>\; =\; 23\; {\rm MeV}, \;\;
  m_{s}<\bar{s}s>\; =\; 215\; {\rm MeV}, \;\;
  m_{h}<\bar{h}h>\; =\; 50\; {\rm MeV}. 
\end{eqnarray} 

\medskip
\begin{eqnarray} \label{eq:set2}
& \mbox{Set 2:}\phantom{setset} & y = 0.50, \; r = 29,\nonumber \\
& & m_{l}<\bar{l}l>\; =\; 30\; {\rm MeV}, \;\;
  m_{s}<\bar{s}s>\; =\; 435\; {\rm MeV}, \;\;
  m_{h}<\bar{h}h>\; =\; 33\; {\rm MeV}.
\end{eqnarray}

\noindent
In Eqs. (\ref{eq:set1}-- \ref{eq:set2}) $l$ stands for light quarks, $s$ is
the strange quark and $h=c,b,t$ denotes heavy quarks. For the light
quarks, we have defined 
$m_{l}<\bar{l}l>$ $\equiv$ 
$\frac{1}{2}[m_u <\bar u u> + m_d <\bar d d>]$.
\noindent
Set 1 and set 2 bracket, at least partially, the present uncertainties. 
In Sect. V.B, in connection with neutralino cosmological properties  
we will also mention the consequences of using a more extreme set of 
values (set 3 of Ref. \cite{noi6}).
For the derivation of the values of the various sets  
see Ref. \cite{noi6}. It is worth noticing that a new derivation of the
pion--nucleon sigma term, $\sigma_{\pi \; N}$, points to rather high values:
$\sigma_{\pi \; N}$ = 73.5$\pm$9 MeV \cite{olsson}. 
By itself, this new result  
would increase the value of the quantity $m_{s}<\bar{s}s>$ given in 
Eq.(\ref{eq:set2}) by $\sim$ 30\%. We recall that the quantity 
$m_{s}<\bar{s}s>$ is crucial in establishing the size of 
$\sigma_{scalar}^{(nucleon)}$ \cite{ggr}. 

As for the values to be assigned to the quantity $\xi = \rho_{\chi}/ \rho_l$ 
we have adopted a standard rescaling recipe. 
For each point of the parameter
space, we take into account the relevant value of the cosmological neutralino
relic density. When $\Omega_\chi h^2$ is larger than a minimal value
$(\Omega h^2)_{\rm min}$, compatible with observational data and with 
large--scale 
structure calculations, we simply put $\xi=1$.
When $\Omega_\chi h^2$ turns out  to be less than $(\Omega h^2)_{\rm min}$, 
and then the neutralino may only provide a fractional contribution
to dark matter, we take $\xi = {\Omega_\chi h^2 / (\Omega h^2)_{\rm min}}$.
The value to be assigned to $(\Omega h^2)_{\rm min}$ is
somewhat arbitrary, in the range 
$0.01 \lsim (\Omega h^2)_{\rm min} \lsim 0.3$. We use here the value 
$(\Omega h^2)_{\rm min} = 0.01$, which is conservatively derived from the
estimate $\Omega_{\rm galactic} \sim 0.03$. 

As we mentioned above, Fig. 1 shows that the annual--modulation region (here 
depicted for $\rho_l$ = 0.3 GeV cm$^{-3}$) is largely covered by the 
scatter plot. This turns out to be the case also for the other 
representative values of $\rho_l$, as is shown in Fig. 2. 
In each panel of this figure we only display the portion of the 
susy scatter plot which is contained in each of the relevant experimental 
region. 
In going from the generic scanning used for Fig. 1 to the one employed for Fig.
2, 
although keeping the overall range of variation of the susy parameter space,
we have optimized the numerical scanning in order to have a number of
configurations, large enough for our subsequent analyses. 
The covering by the scatter plots of the annual--modulation 
regions pertaining to different values of $\rho_l$ is more extended for large 
values of $\rho_l$ than for the small ones, as expected from the features of
the generic plot of Fig. 1.

We define as set $S$ of susy configurations   
the set  comprised of the configurations  whose 
representative points in the plane 
$m_{\chi}$--$\sigma^{(nucleon)}_{scalar}$ lie inside the 
annual--modulation regions displayed in Fig. 2. 
Only configurations of set $S$ are retained in the analyses presented 
hereafter. We remark that set $S$ is the union of all the subsets of susy
configurations which refer to each of the following representative values
for $\rho_l$ and $v_0$:  $\rho_l = 0.1, 0.3, 0.5, 0.7$ GeV cm$^{-3}$,
$v_0 = 170, 220, 270$ km s$^{-1}$, separately.  At any stage, our
results will be analysed and presented in our figures  in terms of the
chosen representative values of  $\rho_l$ and $v_0$, separately.

Another experiment of WIMP direct detection, the CDMS 
experiment \cite{cdms}, is now entering  the 
DAMA sensitivity region. The current CDMS upper bounds (either with or 
without subtractions)  concern the upper left corner
of the annual--modulation 
regions, which is anyway poorly populated by susy configurations; thus, they 
are currently marginal in constraining the set $S$. 
 The upper left corner of the annual--modulation regions is also 
partially disallowed by independent constraints due to indirect measurements 
(see Sect. V).

Now we turn to the constraints which may be applied to the set $S$, 
using data from WIMP indirect search experiments. We set the limits for
exclusion at the same C.L. to which the DAMA region is currently set, i.e. 
99.7\% C.L.

\section{Constraints from neutralino--neutralino annihilation inside 
Earth and Sun}

Indirect evidence for WIMPs in our halo may be obtained at neutrino 
telescopes by measurements of the up--going muons, which would be 
generated by neutrinos produced by pair annihilation of neutralinos 
captured and accumulated inside the Earth and the Sun \cite{flux,mosc,bere2}. 
The size of the expected muon fluxes strongly depends on how these relic 
particles are distributed in the phase space and on the 
intrinsic efficiency of the celestial body in capturing  the 
surrounding WIMPs. 

In the case of the Sun the capture rate is 
essentially determined by its strong  gravitational 
field and by the size of the cross section of neutralino scattering 
off single protons. Instead, in the case of the Earth the capture process 
may quite significantly be enhanced  by coherent neutralino--nucleus 
cross sections, whose size depends on mass--matching condition  between 
$m_{\chi}$ 
and the nuclear mass of the dominant chemical constituents of the 
Earth (O, Si, Mg, Fe) \cite{gould87}.

As for the phase-space neutralino distribution in our neighbourhood, 
together with the usual one based on the standard  Maxwellian 
velocity distribution, whose dispersion speed is centered around 
270 km s$^{-1}$, we also consider two intriguing and conflicting 
models which have been recently discussed in the literature. 

 Damour and Krauss \cite{dk} 
have proposed the existence of a solar--bound population, 
with velocities  restricted to  rather low values,  
$v \lsim 50$ km s$^{-1}$ (for other papers on hypothetical solar-bound WIMP 
populations, see Refs. \cite{steigman,griest,gould89,collar}).   
The Damour--Krauss  solar-bound population
would have been produced by WIMPs which 
scattered off the Sun surface and were set (by perturbations from other 
planets) into orbits which cross the Earth orbit, but not the Sun. 
The ensuing velocities would be distributed in the range 
 25 km s$^{-1} \lsim v \lsim $ 50 km s$^{-1}$.  This population, 
although totally irrelevant for the direct measurements by the DAMA 
NaI--detector, whose electron--equivalent threshold energy is 
2 keV, has been shown to be potentially important in making the 
capture of relic neutralinos by the Earth particularly efficient,  with a 
consequent enhancement of the expected output 
of up--going muons from the Earth, as compared to the standard one 
\cite{bdeku}.  
 For simple kinematical reasons, the 
lower--speed cut off implies that this enhancement is limited to WIMPs of 
masses lower than $\sim$ 150 GeV. 

On the other side, Gould  and Alam \cite{gould99},  using arguments based 
on calculations of asteroids trajectories 
\cite{farinella}, have pointed out
that solar-bound WIMPs could evolve in a way quite different from the one 
derived in Ref. \cite{dk}, with an ensuing  suppression of the 
up-going muon flux usually expected  
from the center of the Earth for a standard halo population. 
This suppression would be significant for  WIMP masses above $\sim$ 65 GeV. 

In the present paper we take into considerations all of these possible 
instances. First we consider the standard situation of a Maxwellian 
velocity distribution over the whole speed range, then we proceed to a 
critical examination of the other two cases, in which the low-speed 
interval  is either overpopulated \cite{dk} or de-populated 
\cite{gould99}, as compared to the standard one.    

The neutrino spectrum and the ensuing up--going muon flux $\Phi_{\mu}$ are
calculated as explained in Refs. \cite{mosc,bere2}. Their normalization 
is set by the annihilation rate $\Gamma_A$ of the neutralinos inside the
celestial body (Earth or Sun), and $\Gamma_A$ depends, in turn, on the capture
rate $C$  of the relic neutralinos by the celestial body 
through the formula 
$\Gamma_A = \frac{C}{2} {\rm tanh}^2 \left (\frac{t}{\tau_A} \right)$ 
\cite{gs}, 
where $t$ is the age of the macroscopic body ($t \simeq 4.5~{\rm Gyr}$ for Sun,
Earth) and 
$\tau_A = (C C_A)^{-1/2}$,  $C_A$ is the annihilation rate 
proportional to the neutralino--neutralino annihilation cross--section and 
$C$ denotes the capture rate. 
In a given macroscopic body the
equilibrium between capture and annihilation ({\it i.e.}
$\Gamma_A \sim C/2$ ) is established
only when $t \gsim \tau_A$. 
Whereas,
in the case of the Sun, the capture--annihilation equilibrium is 
usually reached, due to the much more efficient capture rate due to the
stronger gravitational field, for the Earth, the equilibrium condition is not 
easily realized.

 For the case of the
standard halo population with a Maxwellian velocity distribution, $C$ and
$\Gamma_A$ are calculated as in Refs. \cite{mosc,bere2}, and the ensuing muon
flux is denoted by $(\Phi^{\mathrm Earth}_{\mu})^{\mathrm std}$. 
For the Damour--Krauss population the
quantities $C$ and $\Gamma_A$ are evaluated according to the formulae of Ref.
\cite{bdeku} (the relevant muon flux is denoted by
$(\Phi_{\mu}^{\mathrm Earth})^{\mathrm DK}$). 
For the model conjectured by Gould and Alam \cite{gould99}, we have applied to
the standard capture rate a suppression factor, which we have
re--calculated {\sl ab initio} in the scheme denoted
as {\sl ultra--conservative} in Ref. \cite{gould99}, to cover the whole range
of masses involved in the present paper. 
For many susy configurations the suppression factor in the ensuing up--going
muon fluxes from the center of the Earth is stronger than the reduction
factor in the capture rate alone, due to the relation between $\Gamma_A$ and 
$C$, previously mentioned. 
For these configurations a reduction in the capture
rate induces in the muon flux an extra suppression due to
a critical  increase in the time required for reaching equilibrium. 
 The muon flux calculated in the Gould--Alam model 
is denoted here as $(\Phi_{\mu}^{\mathrm Earth})^{\mathrm GA}$.

All our neutrino fluxes include neutrino oscillations and use the procedure
outlined in Ref. \cite{for}. Here we assume 
$\nu_{\mu} \rightarrow \nu_{\tau}$ oscillations, with values for the 
oscillation parameters 
which are taken from the best fit performed in Ref. \cite{forval} over the whole
set of experimental data on atmospheric neutrinos: 
$\Delta \, m^2 = 3 \cdot 10^{-3}$ eV$^2$, $\sin \theta = 1$. 

Some of our results are presented in Figs. 3--6, where we report various muon
fluxes (or ratios of them) versus $m_{\chi}$, for the four representative
values of $\rho_l$. 
The solid lines, depicted in Fig. 3 and Fig. 6, denote the 99.7\% C.L. upper 
bounds, $(\Phi_{\mu}^{\mathrm Earth})^{lim}$, derived from the data 
 of  the MACRO experiment \cite{macro} 
from the center of the Earth and from the Sun, respectively
(for similar limits from the 
Baksan experiment see Ref. \cite{baksan}).

The scatter plots of Fig. 3 display some expected characteristic features, such
as the peak at $m_{\chi} \sim $ 50--60 GeV, due to the mass--matching between 
$m_{\chi}$ and $m_{\mathrm Fe}$. We notice that a number of configurations induce
up--going muon fluxes in excess of the experimental bounds. Figs. 4--5 show
what would be the enhancement or the reduction effect in
$\Phi_{\mu}^{\mathrm Earth}$
in the case of the Damour--Krauss population or in the Gould--Alam conjecture,
respectively. The size of these effects agree with the evaluations in 
Refs. \cite{bdeku,gould99}. 
For the Damour--Krauss population, the enhancement effect for some susy
configurations appears larger here than in Ref. \cite{bdeku}; this is due to
configurations (not considered in \cite{bdeku}) where rescaling in
$\rho_{\chi}$ is effective. 
In Fig. 6 we display the scatter plots for the up--going
muon flux from the Sun, expected for the standard halo population. The current
experimental bound \cite{macro} sets quite marginal constraints.  

In Sect. V we use the results of this section to constrain the susy
configurations of set $S$. 
The question, as of which model for the low--speed WIMP population among the
two extremes of Refs. \cite{dk,gould99} is applicable, 
is still  open. Thus, 
we implement here the experimental bounds on the standard flux of 
up--going muons; namely, we  exclude 
those configurations, whose 
$(\Phi_{\mu}^{\mathrm Earth})^{\mathrm std}$ is in excess of
the 99.7\% C.L. upper bound derived from the MACRO data.

\section{Constraints from cosmic--ray antiprotons}

The possibility that annihilation of relic particles in the galactic halo 
might distort the spectrum of cosmic--ray antiprotons at low--kinetic energies 
($T_{\bar p} \lsim $1 GeV) has been considered by many authors 
\cite{pbar,cristina,pierre,berg}. 
Indeed, in this energy range, the production of secondary antiprotons by 
interactions of primary cosmic--ray protons with the interstellar 
hydrogen has a kinematical drop off \cite{gaisser}, 
which  primary $\bar p$'s, 
created by relic neutralinos of appropriate mass and composition, might 
 fill in, at least partially.  The effectiveness of this argument to 
disentangle 
ordinary spallation contribution from a possible exotic component due to 
relic particles depends dramatically on how accurately the secondary 
spectrum is calculated \cite{webber,pierre,berg}.  

This point was addressed in Ref. 
\cite{pierre}. In that paper we improved the evaluation of the energy 
losses undergone by secondary antiprotons during their diffusion inside 
 the Galaxy, we noticed that the as--yet most recent  
experimental data (BESS95 \cite{bess95}) were fitted reasonably well 
by the secondary spectrum alone, and we examined critically how much room 
was still available, in the low--energy spectrum, for 
a contribution from an exotic component. Now, 
new experimental data (BESS97 \cite{bess97}) and improved 
evaluations of the secondary spectrum \cite{berg,bieber,ds} further 
constrain the room 
left for  primary sources. These instances, instrumental in 
making the separation between primary and secondary antiprotons more 
difficult,
nevertheless confer to the cosmic--ray antiproton measurements a 
potentially more important 
role in establishing stringent constraints for relic neutralinos of
relatively low mass  in our  halo, once some of the sizeable, still 
persisting, uncertainties are reduced. 

In the present work we have evaluated the primary antiproton flux, 
expected from
neutralino annihilation, as in 
Ref. \cite{pierre}, restricting the supersymmetric configurations to those 
of set $S$. We refer to \cite{pierre} for all
the details concerning the evaluation 
of the production of these primary antiprotons as well as for the 
properties related to their propagation in the halo
and in the heliosphere. Here we only recall 
the features of the neutralino mass distribution function 
adopted in \cite{pierre} as well as here. This mass distribution function 
is taken spheroidal and parameterized 
as a function $\rho_\chi (r,z)$ of the radial distance $r$
from the galactic center in the galactic plane
and of the vertical distance $z$ from the galactic plane in the form 

\begin{equation}
\rho_\chi (r,z) = \rho_{\chi}\,\, \frac{a^2 + r^2_\odot}{a^2 + r^2 +
z^2/f^2},
\label{eq:mass_DF}
\end{equation}

\noindent
where $a$ is the core radius of the halo,
$r_\odot$ is the distance of the Sun from the
galactic center and $f$ is a parameter which describes the flattening  of the
halo.
Here we take the values: $a=3.5$ kpc, $r_\odot = 8$ kpc. 
In the case of a spherical halo ($f = 1$), 
we use the value  $\rho_l = 0.3$ GeV cm$^{-3}$. When $f < 1$ (oblate
spheroidal distribution), $\rho_l$ is taken as \cite{bt,turner1}
\begin{equation}
\rho_l (f) = \rho_l (f = 1)\, \frac{\sqrt{1-f^2}}{f \,
{\rm Arcsin} \sqrt{1-f^2}}. 
\end{equation}

For each value of 
$\rho_l$ and of the relevant value of $f$: 
$\rho_l/({\rm GeV cm}^{-3})$ = 0.1 ($f$ = 1), 0.3  ($f$ = 1), 
0.5 ($f$ = 0.50), 0.7 ($f$ = 0.33), we have evaluated  the top--of--atmosphere (TOA) 
antiproton fluxes,
as the sum of the secondary flux and of 
the primary flux due to neutralino annihilation for the various 
supersymmetric configurations of set $S$, pertaining 
to that specific value of $\rho_l$. The  secondary flux
 has been taken from Ref. \cite{bieber}. 
Re-acceleration effects in the cosmic rays propagation, which might 
also be relevant for the features of the secondary antiproton spectrum 
at low energies \cite{sh,ds}, are not included here. 
Solar modulation has been evaluated according to the procedure discussed in 
Ref. \cite{pierre}. 
We have compared our theoretical results with the combined experimental data 
of BESS95 and BESS97 \cite{bess97}, over the whole experimental energy--range 
(0.18 GeV $\leq T_{\bar p} \leq$ 3.56 GeV), 
by a $\chi^2$ calculation. 

The results are reported in Fig. 7. 
In the evaluation of the $\chi^2$, in addition to the
experimental errors, we have also taken into account 
the  theoretical uncertainties, 
estimated according to the results in Refs. \cite{pierre,berg}, 
with their appropriate energy 
dependence. Orientatively, they are in the following ranges: 
$\pm$ (45--55)\% for the primary fluxes, $\pm$(60--75)\% for the
secondaries, depending on the energy bin. 

In the following, we adopt the selection criterion of excluding 
from set $S$ the configuration whose reduced $\chi^2$ is above the value 
$\chi^2_{r} = 2.44$, which  corresponds to a 99.7\% C.L.   
for the 13 d.o.f. of the BESS 95+97 data.
{}From Fig. 7 we notice that, especially at large values of $\rho_l$, this
constraint disallows a number of susy configurations. The reason why 
the cosmic--ray antiprotons constraint is not more effective 
in constraining set $S$ 
is to be attributed mainly to
the current large uncertainties affecting the evaluation of antiproton
propagation in the galactic halo and in the heliosphere.

\section{Results and conclusions}

Now we apply the experimental bounds from indirect searches
discussed in Sects. III--IV to constrain
the supersymmetric configurations of set $S$. 

\subsection{Combining direct and indirect measurements}

Fig. 8  displays the extent of the covering of the annual--modulation
regions (one for each value of $\rho_l$) by the susy configurations, 
when the MACRO upper bounds are applied to 
$(\Phi_{\mu}^{\mathrm Earth})^{\mathrm std}$. A comparison 
of this figure with Fig. 2 shows that the implementation of these limits
somewhat de-populate the covering regions, with a marked effect for the value 
of the neutralino mass which matches the mass of Iron, as expected. Apart from
this, the extent of the regions covered by the scatter plots does not
significantly change. 

Fig. 9 depicts what would be the
effect for a solar--bound WIMP population {\sl \`a la} Damour--Krauss. 
Especially at low values of $\rho_l$ there would be  some shrinking of the
original regions of the scatter plots in their upper parts, but still the 
annual--modulation regions
would be widely covered by physical susy configurations. 
At variance with this case, the Gould--Alam conjecture would relax the
consequences of the constraints applied in obtaining the plots of Fig. 8.

Now, we return to the case where the experimental bounds 
$(\Phi_{\mu}^{\mathrm Earth})^{lim}$ are applied on  
$(\Phi_{\mu}^{\mathrm Earth})^{\mathrm std}$. 
When, on top of these constraints, we also implement 
the constraints due to cosmic--ray antiprotons, we obtain 
that the scatter plots of Fig. 8 become somewhat de-populated, but without
any appreciable modification in the contours of the covering regions, 
except for a quite marginal downward shift in their upper--left parts. 
Therefore Fig. 8 may be considered as the final situation of our analysis, 
once also the  implementation of the antiprotons constraints has been applied. 
We denote as set $T$ the 
subset of $S$ which comprises the susy configurations not disallowed by 
bounds on the standard up-going muon fluxes and on cosmic--ray antiprotons.

We have analyzed the main properties of the 
configurations of set $T$; some of them are displayed 
 in Figs. 10--11. 
We recall that the scatter plots of these figures 
are derived, as all previous ones, by using for the hadronic quantities, 
discussed in Sect. II, set 1 and set 2, cumulatively. 
In Fig. 10 
we note that the configurations of set $T$ cover  only a specific region 
of the susy parameter space not yet disallowed by accelerator 
constraints. The shape of the distribution of the representative points of 
$T$ in the plot  of Fig. 10 is simply explained by the fact that the values 
of the scalar 
neutralino--nucleon cross section at the level of the DAMA data 
require either a large $\tan \beta$ or a small $m_h$ (or both of these two 
conditions). This constraint is stronger when 
the values of the hadronic quantities are restricted to set 1, alone. 
Fig. 11 
displays a correlation among $m_A$ and $m_0$ which is
mainly due to the interplay of these two quantities in generating
a light $m_h$.
Again, restricting the scatter plot to points belonging to set 1, this 
correlation becomes more pronounced.  
We recall that, at variance with constrained sugra--supersymmetric models, 
 in the MSSM we are using here,   $m_A$ and $m_0$ are treated as independent
 parameters.

\subsection{Cosmological properties}

We turn now to an analysis of the cosmological properties of relic 
neutralinos of the susy configurations of set $T$. 
The relevant plots 
$\Omega_\chi h^2$ vs $m_{\chi}$ are displayed in Fig. 12. 
It is remarkable that the region of main 
cosmological interest:  $\Omega_\chi h^2 \gsim 0.03$ \cite{cosmo} 
turns out to be 
widely populated, with values  of $\Omega_\chi h^2$ which approach, and even
exceed, what may be considered as the current upper 
bound for cold dark matter:
 $\Omega_{CDM} h^2 \lsim $ 0.3 \cite{cosmo}. 
This means that the DAMA annual--modulation data are compatible with a
neutralino as a major component of dark matter. 
We stress that the scatter plot would even shift upward, 
should we use for the hadronic quantities discussed in Sect. II the following 
set: 
$y = 0.50$, $r = 36$, 
$m_{l}<\bar{l}l>$ = 33  MeV, 
$m_{s}<\bar{s}s>$ = 585 MeV, 
$m_{h}<\bar{h}h>$ = 21 MeV. This set of values, denoted as set 3 in Ref. 
\cite{noi6}, is more extreme as compared to set 1 and set 2, but still
compatible with the current uncertainties.  

Finally, we notice that a rather strong de-population in the plots of
Fig. 12 is present around $\Omega_\chi h^2 \simeq 0.01$ and for large
values of $\rho_l$. This effect is induced by the cosmic--ray antiproton
constraint, since the calculated $\bar p$ fluxes have their maximal
values for $\Omega_\chi h^2$ close to the value below which we
apply the rescaling of the local density, i.e. 
$(\Omega h^2)_{\mathrm min} = 0.01$. This property is quite general
in this class of calculations, and it was already commented upon, for
instance, in Ref.\cite{noi}.

\subsection{Conclusions}

In the present paper we have examined the possibility that 
the annual--modulation effect, measured by the DAMA Collaboration 
at a 4$\sigma$ confidence level \cite{dama3}, may be interpreted in terms of 
relic neutralinos. We have examined this problem, by employing the 
Minimal Supersymmetric extension of the Standard Model, as a model 
which does not impose too strong theoretical prejudices on the 
phenomenological analysis.  We have taken into account all 
experimental constraints, from accelerators and from WIMP indirect 
experiments.   

Let us now summarize our main conclusions:

\begin{itemize}

\item The annual--modulation effect mentioned above turns out to be 
      compatible with an interpretation in terms of relic neutralinos.

\item The set of supersymmetric configurations selected by 
      the annual--modulation data is only modestly reduced by current 
      experimental data from WIMP indirect searches (up--going muons from 
      the Earth and the Sun, and cosmic--ray antiprotons).

\item  The set of supersymmetric configurations, selected by 
       the annual--modulation data and not disallowed by the indirect 
       measurements, comprise configurations of relevant cosmological 
       interest, with relic neutralinos playing the role of a major dark 
       matter constituent. 

\end{itemize}

The phenomenological analysis presented in this paper goes beyond 
the discussion of the experimental data specifically discussed here.  
We have tried to pin down the most relevant 
theoretical points, which are still at the origin of large 
uncertainties, and then require additional investigation. 
These are: 
i) size of the  Higgs--quark--quark and the  
squark--quark--neutralino couplings, 
ii) properties of the 
WIMP distribution at low velocities (with the possible existence of a  
solar--bound WIMP population), 
iii) accurate determination of the propagation in the galactic halo and 
in the heliosphere for cosmic--ray antiprotons.

\acknowledgements
This work was partially supported 
by the Research Grants of the Italian Ministero
dell'Universit\`a e della Ricerca Scientifica e Tecnologica 
(MURST) within the {\sl Astroparticle Physics Project}, 
by the Spanish DGICYT under grant number 
PB98--0693, and by the TMR network grant ERBFMRXCT960090 of the European
Union.

\newpage
\begin{center}
{FIGURE CAPTIONS}
\end{center}
\vspace{1cm}

FIG. 1.
Plot of $\xi \, \sigma^{\rm (nucleon)}_{\rm scalar}$ versus $m_{\chi}$. 
The solid line delimits the 
3$\sigma$ C.L.  annual--modulation region, obtained by the DAMA NaI(Tl) 
experiment with a total exposure of 57 986 kg $\cdot$ day \cite{dama3}. 
This region was obtained by including the upper-bound 
constraints of Ref. \cite{dama0}, by setting 
$\rho_l$ at the standard reference value: $\rho_l = 0.3$ GeV cm$^{-3}$, and 
 by taking 
into account uncertainties in the astrophysical velocities of the usual
galactic Maxwellian distribution. 
Also shown in the present figure are the contour
lines for the three values 
$v_0$ = 170 km s$^{-1}$ (short-dashed (red) line), 
$v_0$ = 220 km s$^{-1}$ (long-dash--short-dashed (blue) line), 
$v_0$ = 270 km s$^{-1}$ (long-dashed (green) line), separately. 
The scatter plot is calculated in the MSSM with the scan described in Sect.I;
the points of the scatter plot are coded according to the value of the relic
abundance, $\Omega_{\chi} h^2$, of the relevant susy configuration: dots 
denote $\Omega_{\chi} h^2 < 0.01$, 
crosses denote $0.01 < \Omega_{\chi} \; h^2 < 0.1 $ and 
empty circles denote $\Omega_{\chi} \; h^2 > 0.1$.

\vspace{0.7cm}

FIG. 2.
Location of the DAMA annual--modulation region 
for four  representative values of $\rho_l$: 
$\rho_l$ = 0.1, 0.3, 0.5, 0.7 GeV cm$^{-3}$. 
The scatter plots show only the configurations which lie inside the 
relevant annual--modulation region. A grey--level (color) code 
is used depending on the value of $v_0$ employed in the 
extraction of the annual--modulation region: 
medium grey (red) denotes points which lie in the
annual--modulation region extracted by setting 
$v_0$ = 170 Km s$^{-1}$, 
dark grey (blue) denotes points which lie in the
annual--modulation region extracted by setting 
$v_0$ = 220 Km s$^{-1}$, 
light grey (green) denotes points which lie in the
annual--modulation region extracted by setting 
$v_0$ = 270 Km s$^{-1}$. The three sets are superimposed in that 
sequential order.

\vspace{0.7cm}

FIG. 3.
Scatter plot for the up--going muon flux from the center of the Earth for a
standard Maxwellian distribution, 
$(\Phi_{\mu}^{\mathrm Earth})^{\mathrm std}$,  versus $m_{\chi}$. 
The grey--level (color) code is the same as in Fig. 2. 
The solid line denotes 
the 99.7\% C.L. upper 
bounds, $(\Phi_{\mu}^{\mathrm Earth})^{lim}$, derived from the data  of  the MACRO 
experiment \cite{macro}.

\vspace{0.7cm}

FIG. 4.
Enhancement effect in the up--going muon flux from the center of the Earth 
in case of a solar--bound population {\sl \`a la} Damour--Krauss \cite{dk}. 
The grey--level (color) code is the same as in Fig. 2.

\vspace{0.7cm}

FIG. 5.
Suppression effect in the up--going muon flux from the center of the Earth 
in case of the  Gould--Alam conjecture \cite{gould99}. 
The grey--level (color) code is the same as in Fig. 2. 

\vspace{0.7cm}

FIG. 6.
Scatter plot for the up--going muon flux from the Sun for a
standard Maxwellian distribution, $\Phi_{\mu}^{Sun}$,  versus $m_{\chi}$. 
The grey--level (color) code is the same as in Fig. 2. 
The solid line denotes 
the 99.7\% C.L. upper 
bounds, derived from the data  of  the MACRO 
experiment \cite{macro}. 

\vspace{0.7cm}

FIG. 7.
Scatter plot for the reduced $\chi^2_{r}$'s in a comparison of the calculated 
cosmic--ray antiprotons fluxes with the combined 
experimental 
data of BESS95 and BESS97  \cite{bess97}. The horizontal 
line denotes the value 
$\chi^2_{r}$ = 2.44, which for 13 d.o.f. corresponds to a 
99.7\% C.L., above which we disallow susy configurations. 
The grey--level (color) code is the same as in Fig. 2. 

\vspace{0.7cm}

FIG. 8.
As in Fig. 2, once the constraints from the up--going muon fluxes from the
center of the Earth are applied, assuming a Maxwellian halo distribution for
relic neutralinos.   The grey--level (color) code is the same as in Fig. 2. 

\vspace{0.7cm}

FIG. 9. 
Covering of the annual--modulation regions, if the constraint 
$(\Phi_{\mu}^{\mathrm Earth})^{\mathrm DK} \leq (\Phi_{\mu}^{\mathrm
Earth})^{\mathrm lim}$
were applied. 
The grey--level (color) code is the same as in Fig. 2.

\vspace{0.7cm}

FIG. 10.
Scatter plot for set $T$ in the plane $m_h$ -- $\tan \beta$. 
The grey--level (color) code is the same as in Fig. 2. 
For each panel, the lower dashed line denotes the frontier of the complete 
scatter plot; the upper dashed line denotes the frontier, when only set 1 for
the hadronic quantities of Sect. II is employed. 
The hatched region on the right is excluded by theory. 
The hatched region on the left is 
excluded by present data from LEP \cite{lep2} and CDF \cite{cdf}. 
 The solid line represents the 
95\% C.L. bound reachable at LEP2, in case of non discovery of a neutral 
Higgs boson.

\vspace{0.7cm}

FIG. 11. 
Scatter plot for set $T$ in the plane $m_0$ -- $m_A$. 
The grey--level (color) code is the same as in Fig. 2. 
For each panel, the upper dashed line denotes the frontier of the complete 
scatter plot; the lower dashed line denotes the frontier, when only set 1 for
the hadronic quantities of Sect. II is employed. 

\vspace{0.7cm}

FIG. 12.
Neutralino relic abundance  $\Omega_\chi h^2$ versus $m_{\chi}$, once the
constraints from up--going muon fluxes and cosmic--ray antiprotons are applied. 
The hatched region is disallowed by the upper limit on cold dark matter 
 $\Omega_{CDM} h^2 \lsim $ 0.3 \cite{cosmo}. 

\vspace{0.7cm}


\newpage
\begin{figure}[t]
\hbox{
\psfig{figure=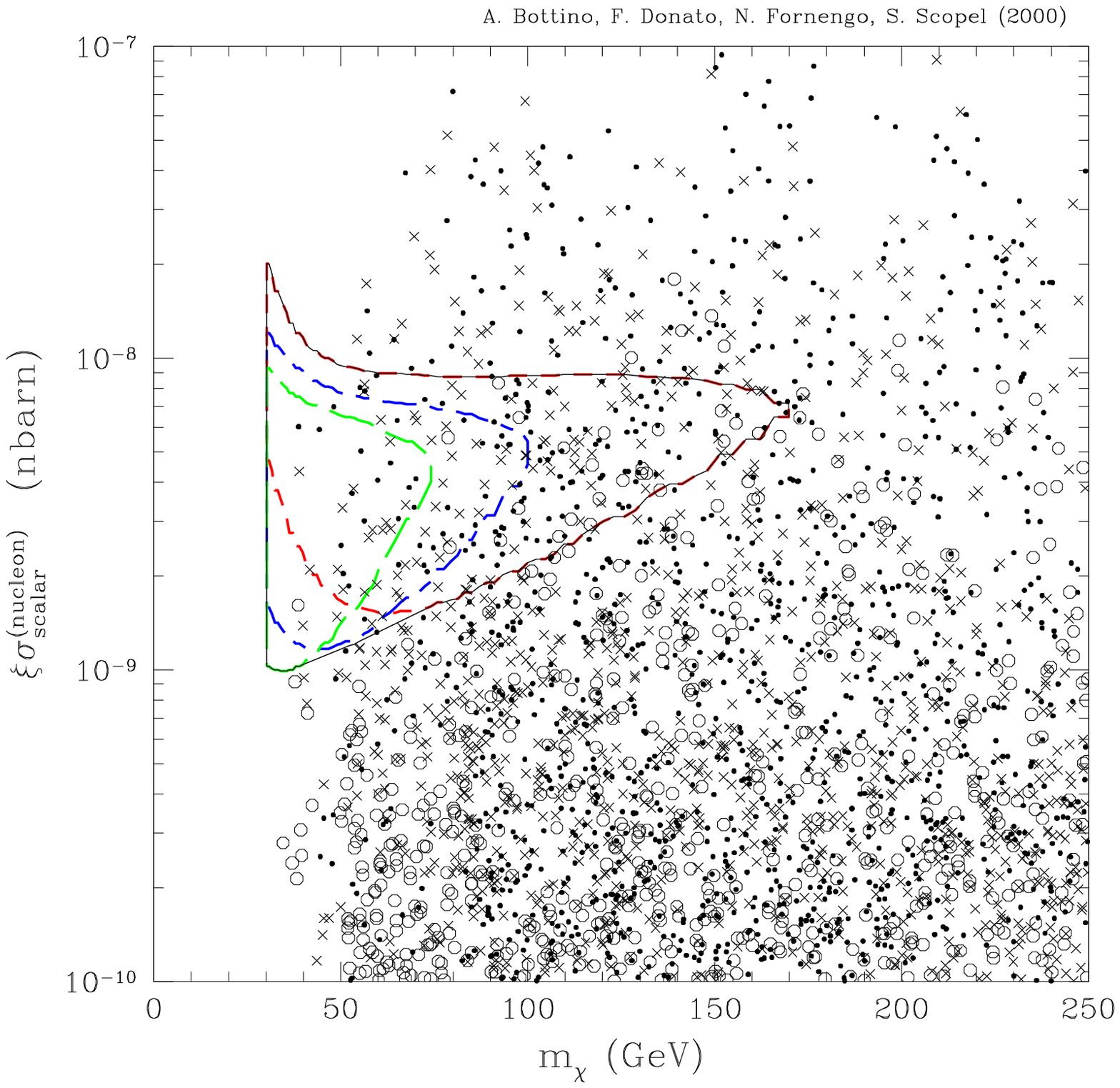,width=8.2in,bbllx=40bp,bblly=160bp,bburx=700bp,bbury=660bp,clip=}
}
{
FIG. 1.
Plot of $\xi \, \sigma^{\rm (nucleon)}_{\rm scalar}$ versus $m_{\chi}$. 
The solid line delimits the 
3$\sigma$ C.L.  annual--modulation region, obtained by the DAMA NaI(Tl) 
experiment with a total exposure of 57 986 kg $\cdot$ day \cite{dama3}. 
This region was obtained by including the upper-bound 
constraints of Ref. \cite{dama0}, by setting 
$\rho_l$ at the standard reference value: $\rho_l = 0.3$ GeV cm$^{-3}$, and 
 by taking 
into account uncertainties in the astrophysical velocities of the usual
galactic Maxwellian distribution. 
Also shown in the present figure are the contour
lines for the three values 
$v_0$ = 170 km s$^{-1}$ (short-dashed (red) line), 
$v_0$ = 220 km s$^{-1}$ (long-dash--short-dashed (blue) line), 
$v_0$ = 270 km s$^{-1}$ (long-dashed (green) line), separately. 
The scatter plot is calculated in the MSSM with the scan described in Sect.I;
the points of the scatter plot are coded according to the value of the relic
abundance, $\Omega_{\chi} h^2$, of the relevant susy configuration: dots 
denote $\Omega_{\chi} h^2 < 0.01$, 
crosses denote $0.01 < \Omega_{\chi} \; h^2 < 0.1 $ and 
empty circles denote $\Omega_{\chi} \; h^2 > 0.1$. 
}
\end{figure}

\newpage
\begin{figure}[t]
\hbox{
\psfig{figure=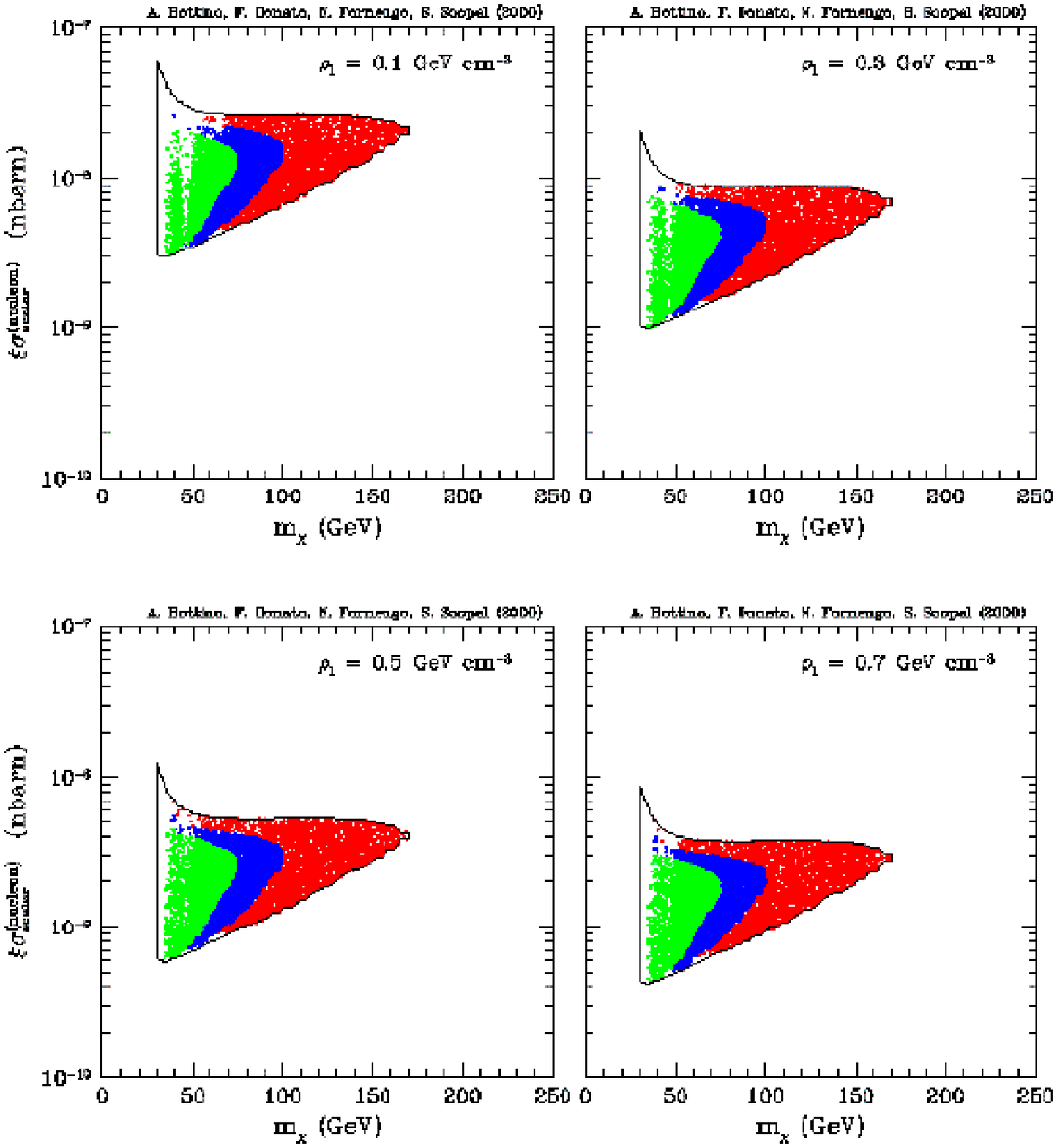,width=7.5in,bbllx=27bp,bblly=100bp,bburx=700bp,bbury=717bp,clip=}
}
{
FIG. 2.
Location of the DAMA annual--modulation region 
for four  representative values of $\rho_l$: 
$\rho_l$ = 0.1, 0.3, 0.5, 0.7 GeV cm$^{-3}$. 
The scatter plots show only the configurations which lie inside the 
relevant annual--modulation region. A grey--level (color) code 
is used depending on the value of $v_0$ employed in the 
extraction of the annual--modulation region: 
medium grey (red) denotes points which lie in the
annual--modulation region extracted by setting 
$v_0$ = 170 Km s$^{-1}$, 
dark grey (blue) denotes points which lie in the
annual--modulation region extracted by setting 
$v_0$ = 220 Km s$^{-1}$, 
light grey (green) denotes points which lie in the
annual--modulation region extracted by setting 
$v_0$ = 270 Km s$^{-1}$. The three sets are superimposed in that 
sequential order. 
}
\end{figure}

\newpage
\begin{figure}[t]
\hbox{
\psfig{figure=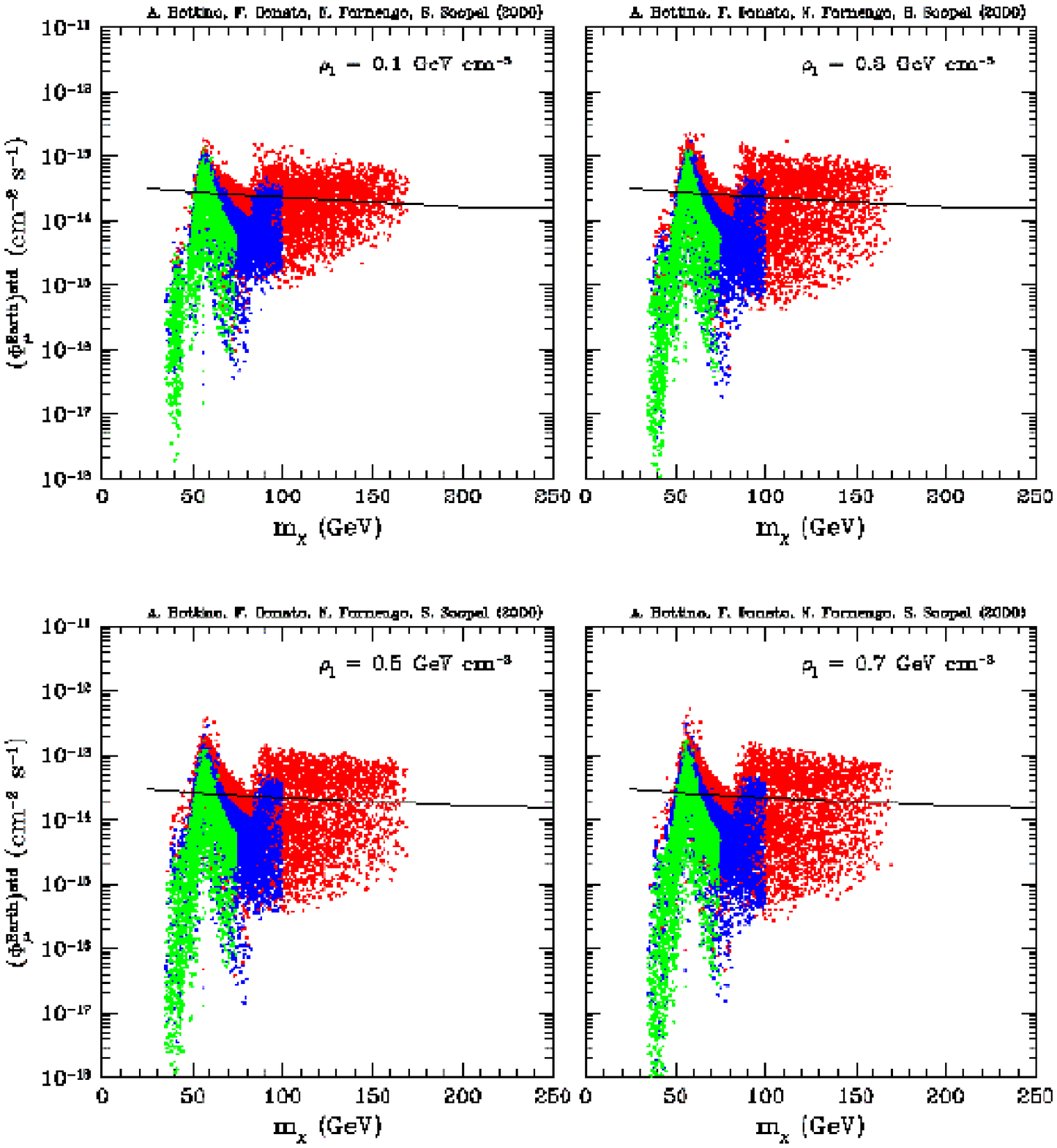,width=7.5in,bbllx=27bp,bblly=100bp,bburx=700bp,bbury=717bp,clip=}
}
{
FIG. 3.
Scatter plot for the up--going muon flux from the center of the Earth for a
standard Maxwellian distribution, 
$(\Phi_{\mu}^{\mathrm Earth})^{\mathrm std}$,  versus $m_{\chi}$. 
The grey--level (color) code is the same as in Fig. 2. 
The solid line denotes 
the 99.7\% C.L. upper 
bounds, $(\Phi_{\mu}^{\mathrm Earth})^{lim}$, derived from the data  of  the MACRO 
experiment \cite{macro}. 
}
\end{figure}

\newpage
\begin{figure}[t]
\hbox{
\psfig{figure=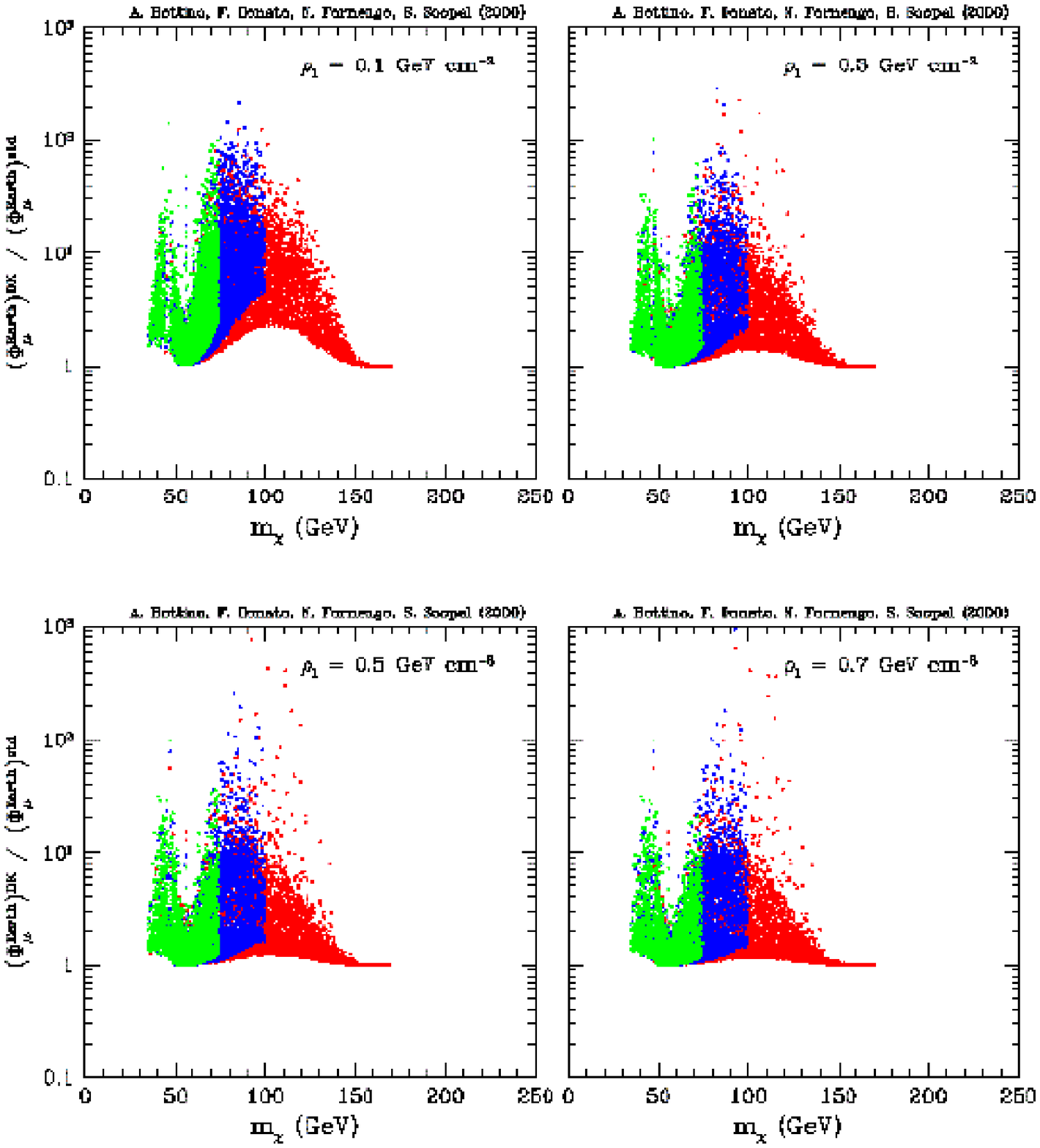,width=7.5in,bbllx=27bp,bblly=100bp,bburx=700bp,bbury=717bp,clip=}
}
{
FIG. 4.
Enhancement effect in the up--going muon flux from the center of the Earth 
in case of a solar--bound population {\sl \`a la} Damour--Krauss \cite{dk}. 
The grey--level (color) code is the same as in Fig. 2. 
}
\end{figure}

\newpage
\begin{figure}[t]
\hbox{
\psfig{figure=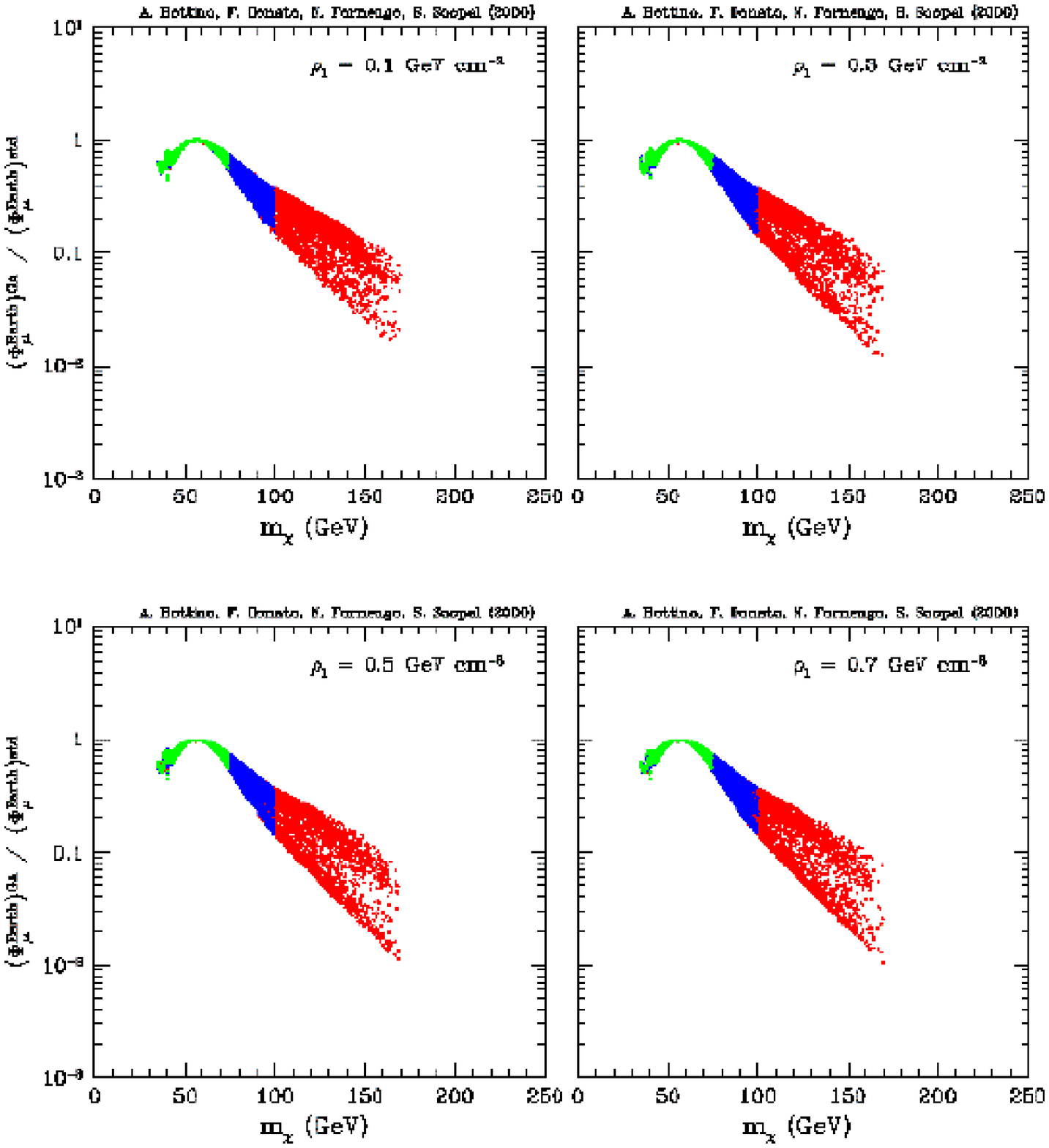,width=7.5in,bbllx=27bp,bblly=100bp,bburx=700bp,bbury=717bp,clip=}
}
{
FIG. 5.
Suppression effect in the up--going muon flux from the center of the Earth 
in case of the  Gould--Alam conjecture \cite{gould99}. 
The grey--level (color) code is the same as in Fig. 2. 
}
\end{figure}

\newpage
\begin{figure}[t]
\hbox{
\psfig{figure=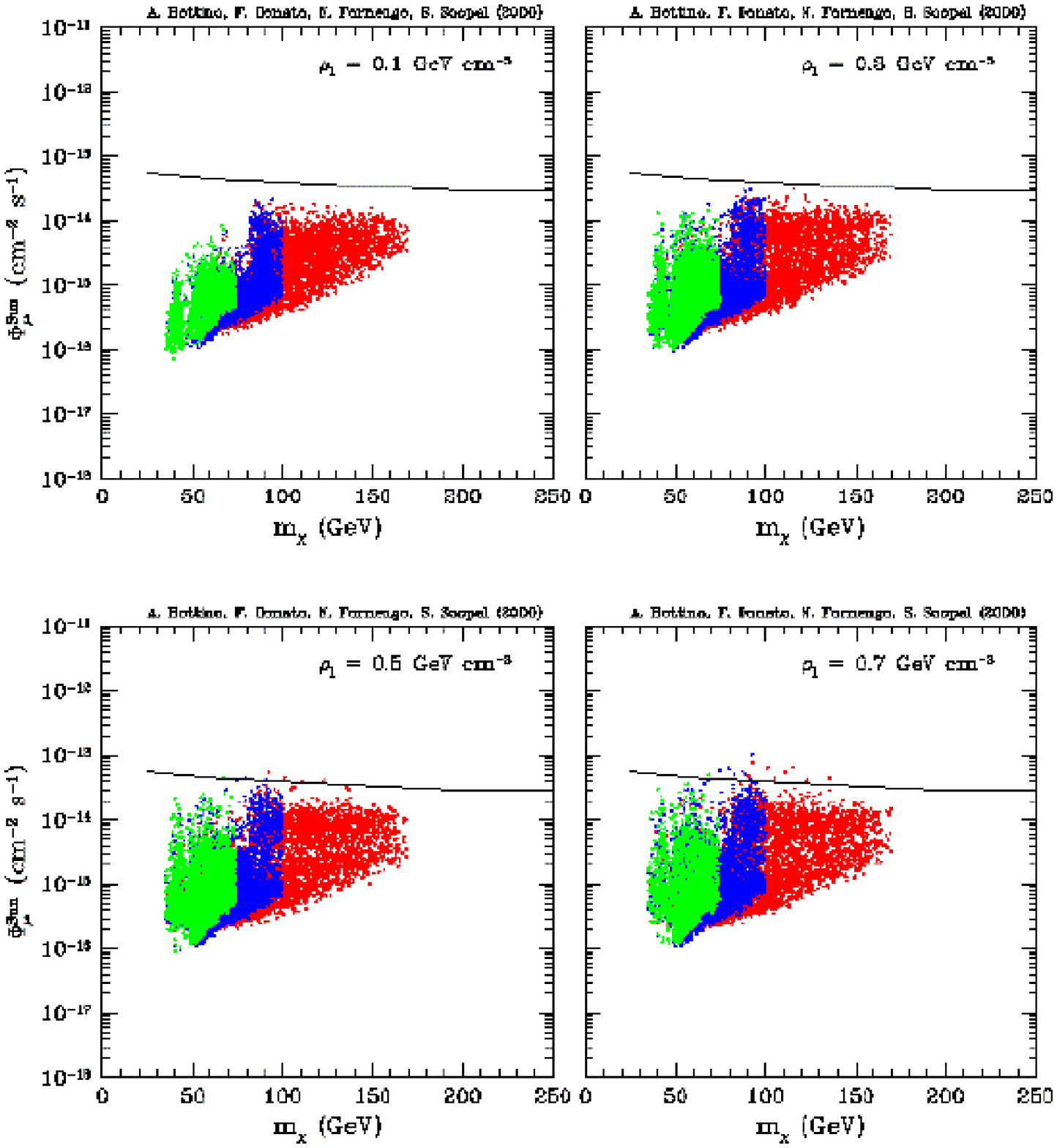,width=7.5in,bbllx=27bp,bblly=100bp,bburx=700bp,bbury=717bp,clip=}
}
{
FIG. 6.
Scatter plot for the up--going muon flux from the Sun for a
standard Maxwellian distribution, $\Phi_{\mu}^{Sun}$,  versus $m_{\chi}$. 
The grey--level (color) code is the same as in Fig. 2. 
The solid line denotes 
the 99.7\% C.L. upper 
bounds, derived from the data  of  the MACRO 
experiment \cite{macro}. 
}
\end{figure}

\newpage
\begin{figure}[t]
\hbox{
\psfig{figure=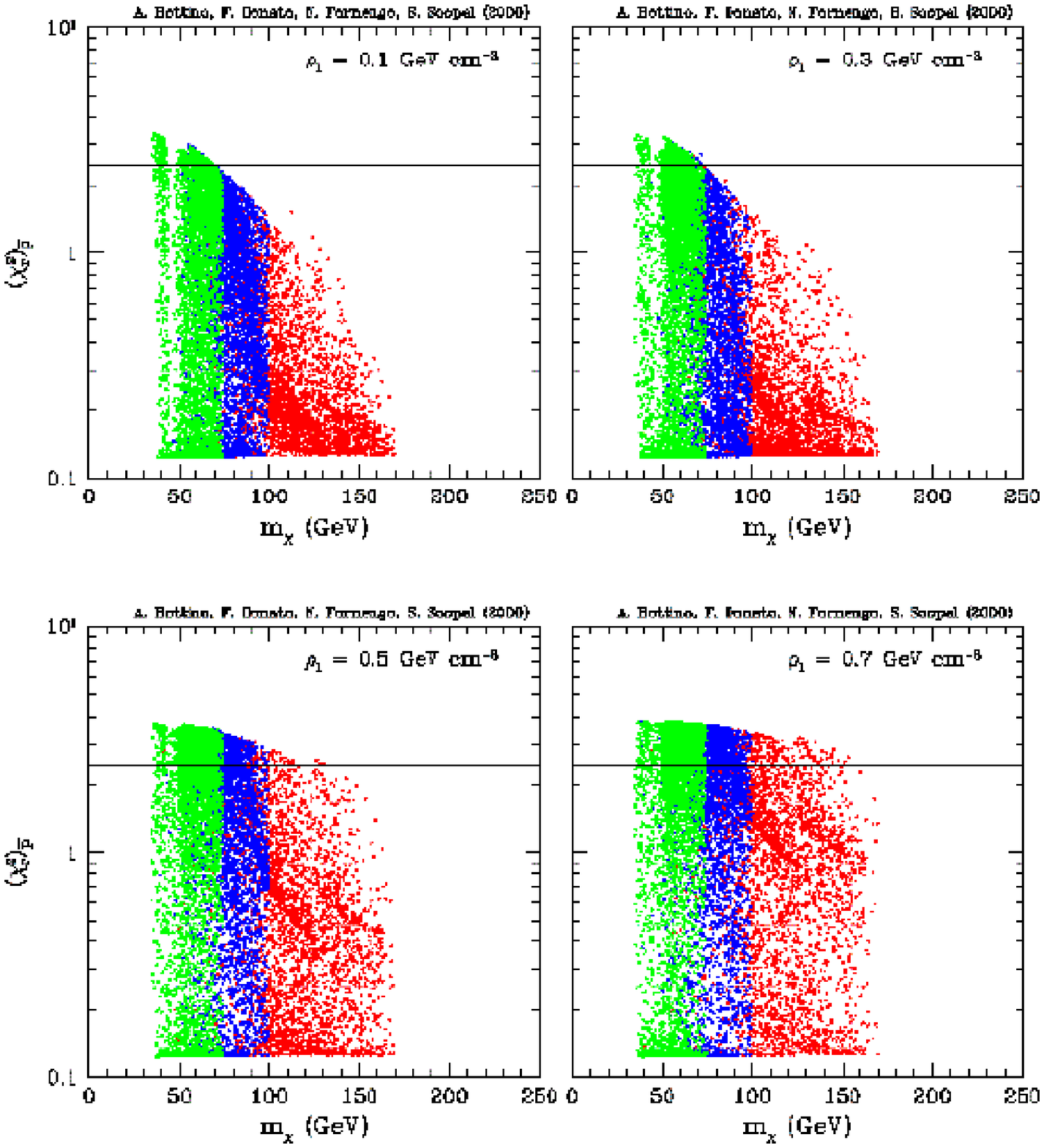,width=7.5in,bbllx=27bp,bblly=100bp,bburx=700bp,bbury=717bp,clip=}
}
{
FIG. 7.
Scatter plot for the reduced $\chi^2_{r}$'s in a comparison of the calculated 
cosmic--ray antiprotons fluxes with the combined 
experimental 
data of BESS95 and BESS97  \cite{bess97}. The horizontal 
line denotes the value 
$\chi^2_{r}$ = 2.44, which for 13 d.o.f. corresponds to a 
99.7\% C.L., above which we disallow susy configurations. 
The grey--level (color) code is the same as in Fig. 2. 
}
\end{figure}

\newpage
\begin{figure}[t]
\hbox{
\psfig{figure=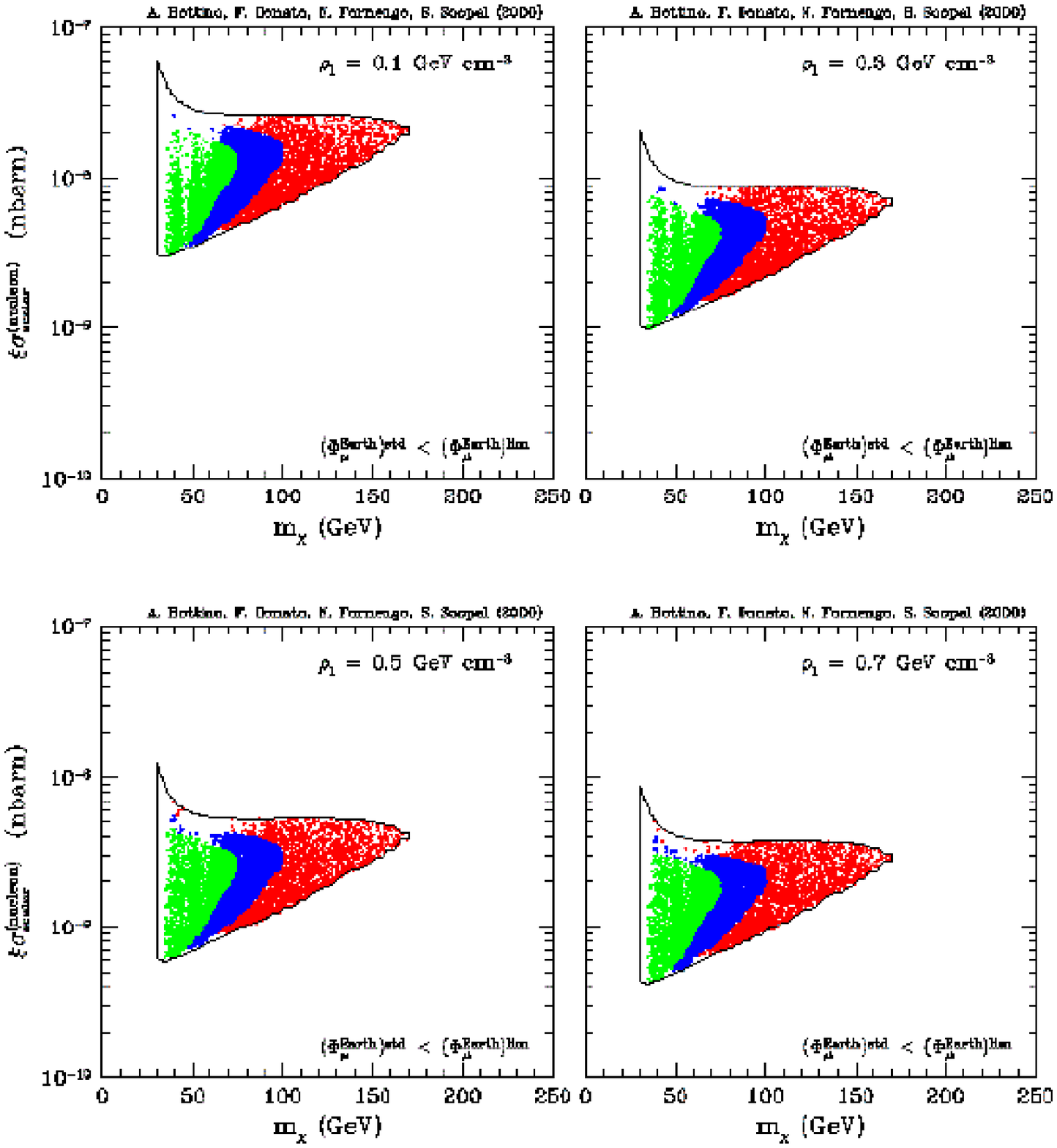,width=7.5in,bbllx=27bp,bblly=100bp,bburx=700bp,bbury=717bp,clip=}
}
{
FIG. 8.
As in Fig. 2, once the constraints from the up--going muon fluxes from the
center of the Earth are applied, assuming a Maxwellian halo distribution for
relic neutralinos.   The grey--level (color) code is the same as in Fig. 2. 
}
\end{figure}

\newpage
\begin{figure}[t]
\hbox{
\psfig{figure=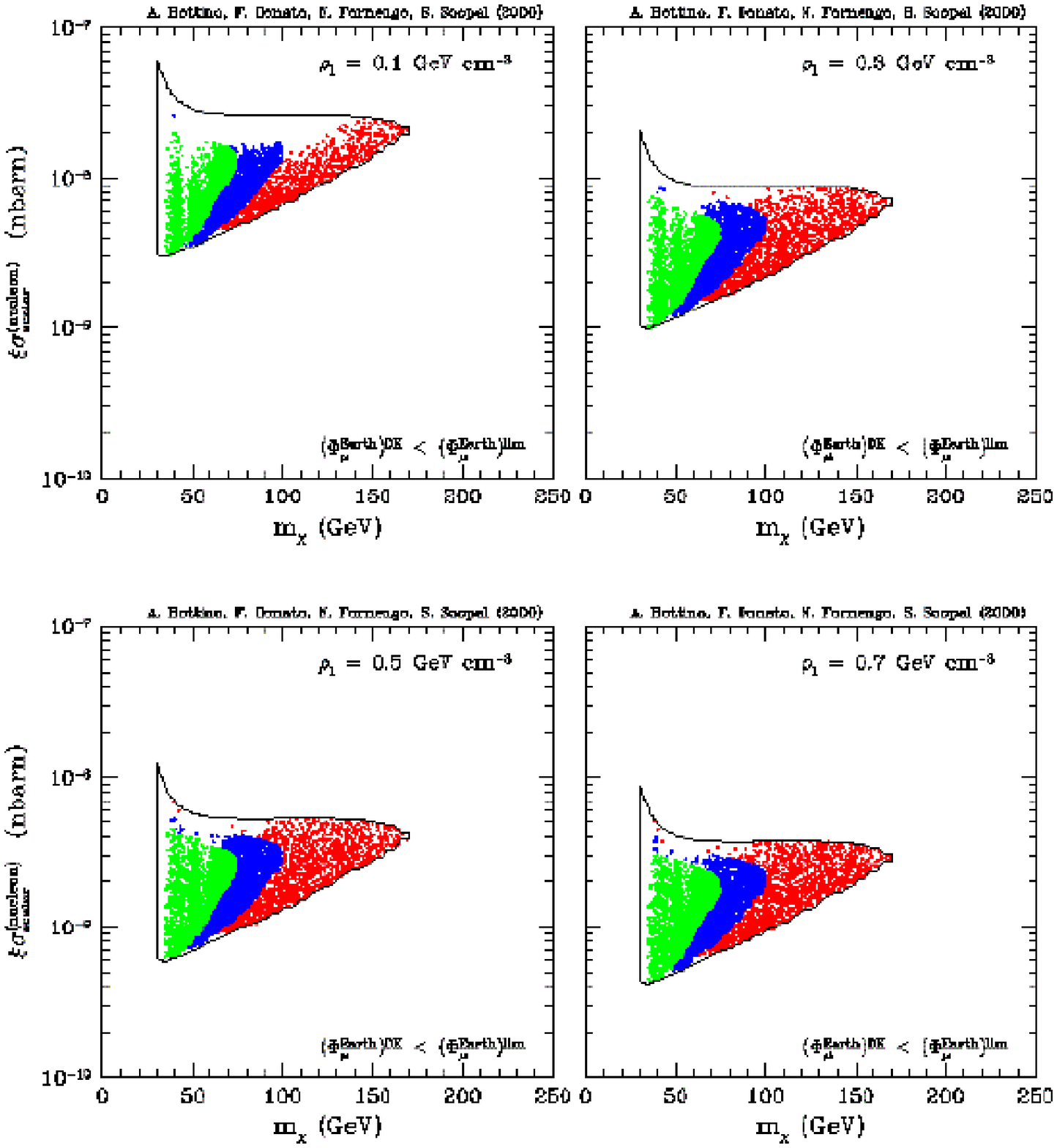,width=7.5in,bbllx=27bp,bblly=100bp,bburx=700bp,bbury=717bp,clip=}
}
{
FIG. 9. 
Covering of the annual--modulation regions, if the constraint 
$(\Phi_{\mu}^{\mathrm Earth})^{\mathrm DK} \leq (\Phi_{\mu}^{\mathrm
Earth})^{\mathrm lim}$
were applied. 
The grey--level (color) code is the same as in Fig. 2. 
}
\end{figure}

\newpage
\begin{figure}[t]
\hbox{
\psfig{figure=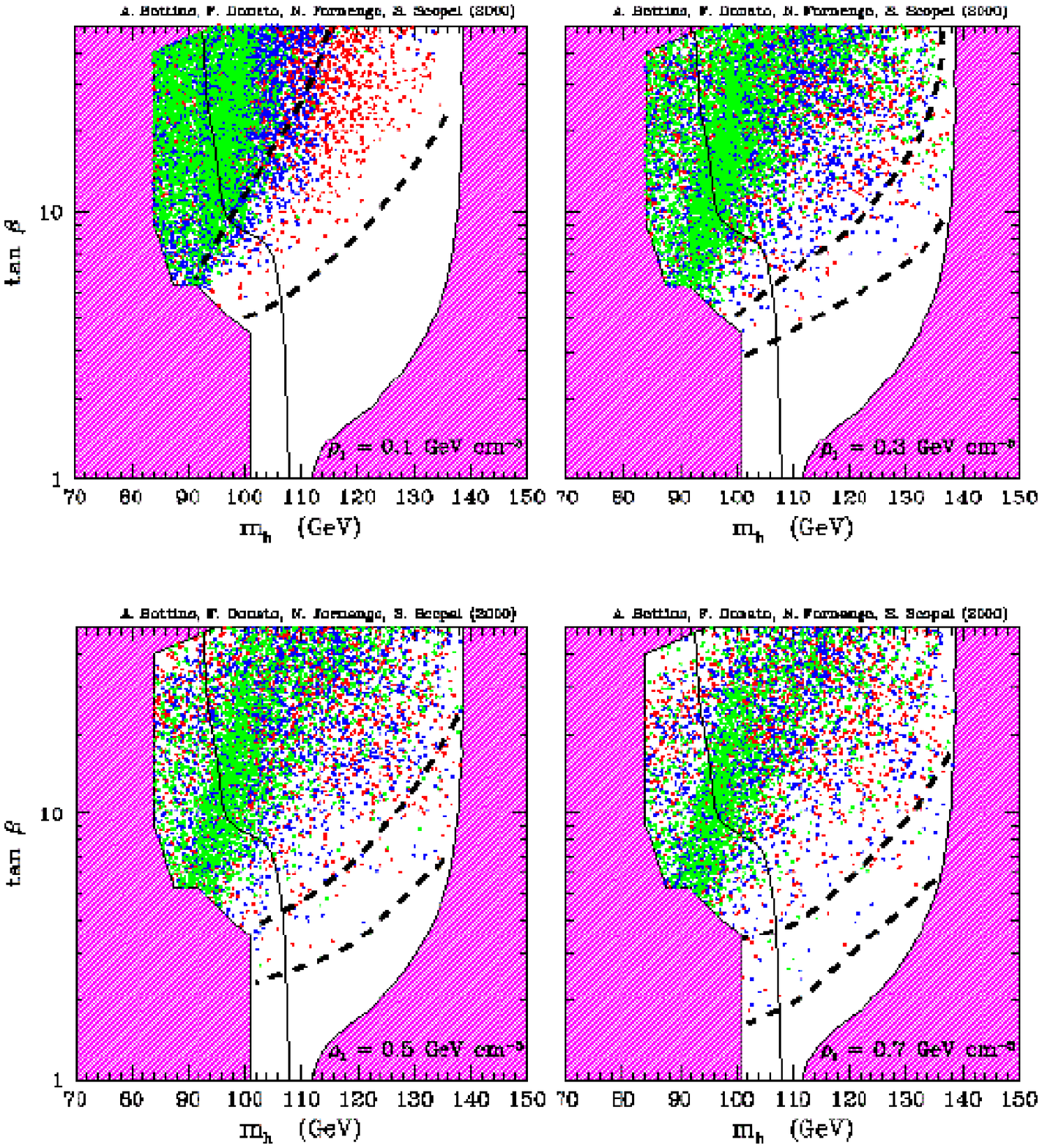,width=7.5in,bbllx=27bp,bblly=100bp,bburx=700bp,bbury=717bp,clip=}
}
{
FIG. 10.
Scatter plot for set $T$ in the plane $m_h$ -- $\tan \beta$. 
The grey--level (color) code is the same as in Fig. 2. 
For each panel, the lower dashed line denotes the frontier of the complete 
scatter plot; the upper dashed line denotes the frontier, when only set 1 for
the hadronic quantities of Sect. II is employed. 
The hatched region on the right is excluded by theory. 
The hatched region on the left is 
excluded by present data from LEP \cite{lep2} and CDF \cite{cdf}. 
 The solid line represents the 
95\% C.L. bound reachable at LEP2, in case of non discovery of a neutral 
Higgs boson. 
}
\end{figure}

\newpage
\begin{figure}[t]
\hbox{
\psfig{figure=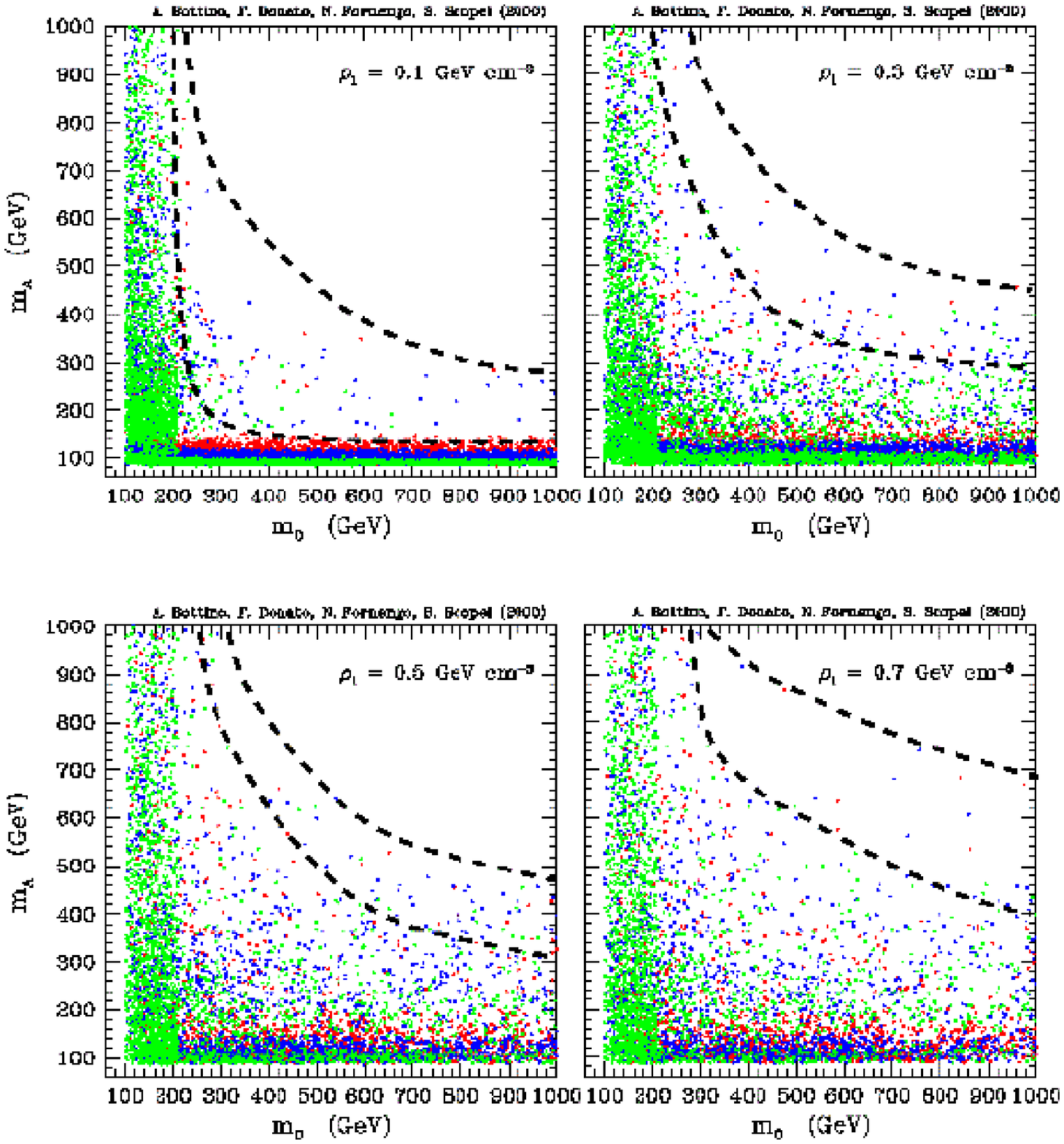,width=7.5in,bbllx=27bp,bblly=100bp,bburx=700bp,bbury=717bp,clip=}
}
{
FIG. 11. 
Scatter plot for set $T$ in the plane $m_0$ -- $m_A$. 
The grey--level (color) code is the same as in Fig. 2. 
For each panel, the upper dashed line denotes the frontier of the complete 
scatter plot; the lower dashed line denotes the frontier, when only set 1 for
the hadronic quantities of Sect. II is employed. 
}
\end{figure}

\newpage
\begin{figure}[t]
\hbox{
\psfig{figure=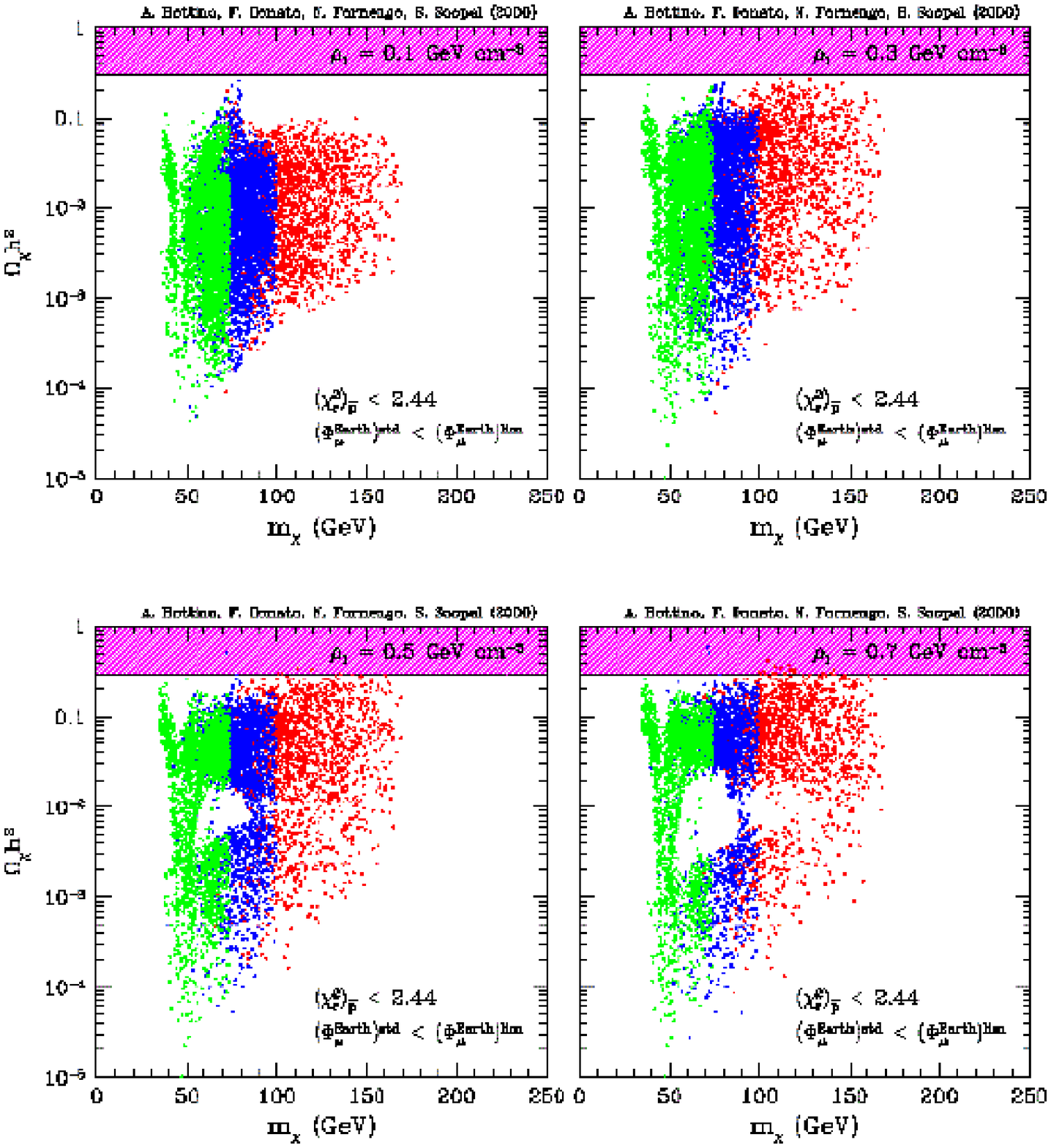,width=7.5in,bbllx=27bp,bblly=100bp,bburx=700bp,bbury=717bp,clip=}
}
{
FIG. 12.
Neutralino relic abundance  $\Omega_\chi h^2$ versus $m_{\chi}$, once the
constraints from up--going muon fluxes and cosmic--ray antiprotons are applied. 
The hatched region is disallowed by the upper limit on cold dark matter 
 $\Omega_{CDM} h^2 \lsim $ 0.3 \cite{cosmo}. 
}
\end{figure}

\end{document}